\documentclass[aps,prl,twocolumn,superscriptaddress,longbibliography]{revtex4-1}
\usepackage{graphicx} 
\usepackage{float} 
\usepackage{dcolumn} 
\usepackage{bm,color}
\usepackage{amssymb,dsfont,amstext,amsfonts}
\usepackage{amsmath}
\usepackage{extarrows}
\usepackage[colorlinks=true,linkcolor=blue,urlcolor=blue,citecolor=blue]{hyperref}
\usepackage{xcolor}
\usepackage{wasysym}
\usepackage{mathtools}
\usepackage{bbold}

\usepackage[normalem]{ulem} 
\usepackage{color}

\newcommand{\la}{\langle}
\newcommand{\ra}{\rangle}
\renewcommand{\ao}{\hat{a}^{\phantom{\dag}}}
\renewcommand{\aa}{\hat{a}^\dag}
\newcommand{\Ho}{\hat{H}}
\newcommand{\Uo}{\hat{U}}
\newcommand{\bk}{\bm{k}}
\newcommand\vex[1]{\bm{#1}}
\newcommand\gvex[1]{\bm{#1}}

\def\sgn{\mathrm{sgn}}

\def\dd{\mathrm{d}}

\def\nodag{^{\vphantom{\dagger}}}

\def\mod{\:\mathrm{mod}\:}

\begin{document}
\title{How to Directly Measure Floquet Topological Invariants in Optical Lattices}

\author{F. Nur \"Unal}
\email{unal@pks.mpg.de}
\affiliation{Max Planck Institute for the Physics of Complex Systems, N\"othnitzer Stra\ss e 38, Dresden 01187, Germany}
\author{Babak Seradjeh}
\email{babaks@indiana.edu}
\affiliation{Max Planck Institute for the Physics of Complex Systems, N\"othnitzer Stra\ss e 38, Dresden 01187, Germany}
\affiliation{Department of Physics, Indiana University, 727 E Third Street, Bloomington, Indiana 47405, USA}
\author{Andr\'e Eckardt}
\email{eckardt@pks.mpg.de}
\affiliation{Max Planck Institute for the Physics of Complex Systems, N\"othnitzer Stra\ss e 38, Dresden 01187, Germany}

\date{\today}

\begin{abstract}
The classification of topological Floquet systems with time-periodic Hamiltonians transcends that of static systems. For example, spinless
fermions in periodically driven two-dimensional lattices are not completely characterized by the Chern
numbers of the quasienergy bands, but rather by a set of winding numbers associated with the gaps. We propose a feasible scheme for measuring these winding numbers in a periodically driven optical lattice efficiently and directly. It is based on the construction of a one-parameter family of drives, continuously connecting the Floquet system of interest to a trivial reference system. The winding numbers are then determined by the identification and the tomography of the band-touching singularities occurring on the way. As a byproduct, we also propose a method for probing spectral properties of time evolution operators via a time analog of crystallography.
\end{abstract}

\maketitle

\textit{Introduction}.---%
Topological insulators are characterized by quantized topological invariants. These are nonlocal bulk
properties distinguishing different gapped phases, which change only at energy gap closings, and determine robust properties of the system~\cite{HasanKane10_Rev,Xiao11_Rev}. For instance, Chern numbers describing generic two-dimensional (2D) band insulators dictate the number of chiral edge modes at the system's boundary and quantize the Hall response. The Chern number can be inferred from measuring the Hall response~\cite{Klitzing80_PRL,Aidelsburger15_NatPhys}, the circular dichroism~\cite{Asteria18_arx}, by observing chiral edge transport~\cite{Mancini15_Sci,Stuhl15_Sci,Rechtsman13_Nat} or via interferometry \cite{Duca15_Science,Atala13_NatPhys}.  However, a direct measurement of the Chern number characterizing a topologically non-trivial band structure was achieved only very recently~\cite{Tarnowski17_arx,Flurin17_Science}.

In recent years it was shown that periodically driven (Floquet) systems can also possess robust topological properties~\cite{OkaAok09a,JiaKitAli11a,Eckardt17_Rev, KunSer13a,KunFerSer14a, AniZlaAnd15a,KunFerSer16a,Wang18_PRL,Raciunas18_PRA,RacZlaEck16a,RodFerSer18a,RodSer18a,Kolovsky11_EPL,Bermudez11_PRL,Hauke12_PRL, Jotzu14_Nat,Aidelsburger13_PRL,Miyake13_PRL,Rechtsman13_Nat,HuPillay15_PRX,GaoGao16_NatCommun,Mukherjee17_NatComm, Mukherjee18_NatComm,Maczewsky17_NatCommun} in ways that transcend static systems~\cite{Kitagawa10_PRB,Rudner13_PRX,Nathan15_NJP}. While the classification of Floquet topological phases has been established~\cite{RoyHar17a,YaoYanWan17a} and corresponding anomalous edge modes been observed in photonic systems~\cite{HuPillay15_PRX,GaoGao16_NatCommun,Mukherjee17_NatComm,Mukherjee18_NatComm,Maczewsky17_NatCommun}, to date a direct measurement of their topological invariants or even a scheme for doing so is lacking. In this Letter, we describe how such a direct measurement of Floquet topological invariants can be performed in fermionic optical-lattice systems.

Consider a generic 2D lattice under time-periodic modulation. The quasistationary Floquet-Bloch states and their quasienergy bands $\varepsilon_b(\bk)$, indexed by $b$ and quasimomentum wave vectors
$\bk$, can be obtained from diagonalizing the effective time-independent Floquet Hamiltonian, which generates the stroboscopic evolution of
the system in steps of the driving period $T$. Like in equilibrium, the quasienergy bands can be assigned Chern numbers $C_b$. However, these Chern numbers are not sufficient for a complete
characterization of the topological properties the Floquet system~\cite{Kitagawa10_PRB}.
The distinction becomes apparent when considering the bulk-boundary correspondence~\cite{Rudner13_PRX}. Since the quasienergies are defined modulo the drive frequency $\omega=2\pi/T$ only, they can be chosen to lie in the Floquet Brillouin zone (FBZ) $T\varepsilon_b \in (-\pi,\pi]$
(hereafter $\hbar=1$). Defined on a ring, the $N$ quasienergy bands are separated
by $N$ gaps, for $N$ basis points in the unit cell.
Each quasienergy band gap $g$ is assigned an integer-valued invariant, $W_g$, obtained from the spatiotemporal winding structure of the full time-evolution operator~\cite{Rudner13_PRX}. For a system with an open boundary, the $N$ independent Floquet topological invariants $W_g$ count the number of topologically protected chiral edge modes crossing gap $g$. Denoting the gap above (below) a band $b$ with ${>_b}$ (${<_b}$), one finds $C_b = W_{>_b}-W_{<_b}$~\cite{Rudner13_PRX}. As a result, Chern numbers add up to zero when summed over all bands and are described by $N-1$ independent integers only.
Thus, only for static systems characterized by $N-1$ energy gaps Chern numbers provide a full characterization.

In the following, we propose two schemes for directly  measuring the winding numbers $W_g$ in a system of fermionic atoms in a driven optical lattice: A brute-force approach based on the tomography of the full one-cycle evolution operator as well as a much more efficient scheme. The latter relies on the construction of a one-parameter family of drives connecting the Floquet system to a trivial reference point and the identification and the tomography of band-touching singularities occurring in this family.

\begin{figure}
	\centering\includegraphics[width=1\linewidth]{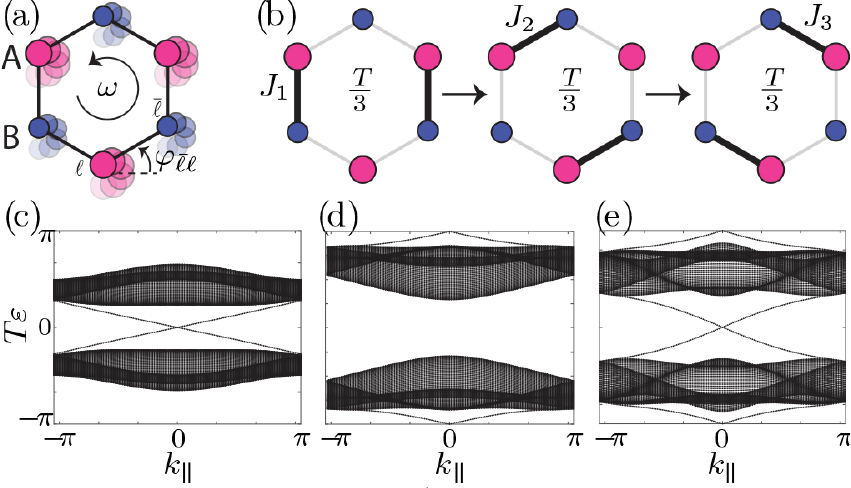}
	\caption{Driven honeycomb lattice. (a) Continuous drive via circular shaking. (b) Step-wise drive. (c-e) Floquet spectra for continuously driven model on a finite strip with armchair termination, for the quasimomentum $k_\parallel$ along periodic direction along the strip for $\nu=1$. The parameters are (c)
	$\omega=3J$, $\delta=-2J$, $\kappa=2$; (d) $\omega=4J$, $\delta=0.2J$, $\kappa=1$; (e) $\omega=2.5J$,
	$\delta=-2J$, $\kappa=1.5$. }
    \label{fig - edgeSpectrum}
\end{figure}

\textit{The system}.---%
We consider a driven two-dimensional system of spin-polarized fermions
in an optical lattice with $N=2$ sublattice states per unit cell. It is described by
a time-periodic Hamiltonian $\Ho(t)=\Ho(t+T)$ and the time-evolution operator
$\Uo(t)=\mathcal{T}\exp[-i\int_0^t\Ho(t')\dd t']$ with time ordering $\mathcal{T}$. Thanks to the discrete time
translation invariance, the stroboscopic
evolution of the system over each drive cycle can be described by the time-independent
Floquet Hamiltonian $\Ho_F=i \log[\Uo(T)]/T$, whose eigenvalues define the quasienergy spectrum~\cite{Eckardt17_Rev}. The branch of the natural logarithm shall be chosen so that the single-particle quasienergies lie in the FBZ.
The winding numbers of the system cannot be extracted solely from the Floquet Hamiltonian; they contain
information also about the micromotion captured by the time-periodic operator
$\Uo_F(t)=\Uo(t)e^{iH_Ft}$.

The Hamiltonian takes the tight-binding form
\begin{equation}
\Ho(t)=-\sum_{\la\bar\ell\ell\ra}J_{\bar\ell\ell}(t)\aa_{\bar\ell}\ao_\ell
				+\frac{\Delta}{2} \sum_\ell\eta_\ell \aa_\ell\ao_\ell,
\end{equation}
where $\ao_\ell$ annihilates a fermion on lattice site $\ell$ of sublattice $\mathsf{A}$ or $\mathsf{B}$, for which $\eta_\ell=+1$ and $-1$, respectively. They are
separated by the energy offset $\Delta=\nu\omega + \delta$, with the resonant part
$\nu\omega$ and detuning $\delta$, where the integer $\nu$ is chosen so that $|\delta|\le \omega/2$. The time dependency is encoded in the tunneling matrix elements between neighboring sites $J_{\bar\ell\ell}(t)$, which also describe time-periodic
conservative forces via modulated Peierls phases $\arg[J_{\bar\ell\ell}(t)]$.

For periodic boundary conditions, the Hamiltonian becomes diagonal with respect to quasimomentum $\bk$,
\begin{equation}
\hat H(t)= \sum_{\vex k} \hat a_{\vex k}^\dagger {\cal H}(\vex k,t) \hat a\nodag_{\vex k}.
\end{equation}
Here $\hat a_{\vex k}^\dagger = (\hat a_{\mathsf{A}\vex k}^\dagger, \hat a_{\mathsf{B}\vex k}^\dagger)$ is the spinor of fermionic creation operators defined by $\aa_{\mathsf{s}\bk}=\frac1{\sqrt{M}}\sum_{\ell\in \mathsf{s}}e^{i\bk\cdot\bm{r}_\ell} \aa_\ell$, with $\mathsf{s}=\mathsf{A},\mathsf{B}$, number of unit cells $M$, and site positions $\bm{r}_\ell$. The single-particle Hamiltonian ${\cal H}(\bm{k},t)$ can be expressed in terms of the vector of Pauli matrices $\bm{\sigma}$ acting in sublattice space, ${\cal H}(\bm{k},t)= \bm{h}(\bm{k},t)\cdot\gvex{\sigma}$. Likewise, we  express the time-evolution $\mathcal{U}(\bk,t)$ and the Floquet Hamiltonian $\mathcal{H}_F(\bk)=\bm{h}_F(\bk)\cdot\gvex{\sigma}$ in momentum space. The two quasienergy bands $\varepsilon_\pm(\bk)=\pm|\bm h_F(\bk)|$ are separated by two gaps $\gamma_g$, which we label by $g=0$ and $\pi$: $\gamma_0(\bk)=\varepsilon_+(\bk)-\varepsilon_-(\bk)=2|\bm h_F(\bk)|$ is centered around the FBZ center, $T\varepsilon=0$, and $\gamma_\pi(\bk)=\omega -\gamma_0(\bk)$ is centered around the zone edge, $T\varepsilon=\pi$.

In order to illustrate our results, we consider two concrete model systems defined on a hexagonal lattice. The first one is driven by a continuous circular force, which can be realized by lattice shaking \cite{EckardtEtAl10} [cf.\ Fig.~\ref{fig - edgeSpectrum}(a)]. It describes recent quantum-gas experiments, where topologically non-trivial band structures were engineered via periodic driving \cite{Jotzu14_Nat,Tarnowski17_arx,Flaschner16_Sci,Flaschner18_Nat,Asteria18_arx}. The force is captured by time-periodic Peierls phases, $J_{\bar\ell\ell}(t)=J e^{i\kappa\sin(\omega t-\varphi_{\bar\ell\ell})}$, with drive strength $\kappa$ and polar angle $\varphi_{\bar\ell\ell}$ denoting the direction of tunneling. We further assume $\nu=1$, which is naturally suitable for state tomography~\cite{Flaschner18_Nat,Tarnowski17_arx}. The second model was introduced in Ref.~\cite{Kitagawa10_PRB} with $\nu=0$. It is driven in three steps of equal duration $T/3$, during each of which the tunneling parameters take a non-zero value $J$ only along one of the three possible directions [cf.\ Fig.~\ref{fig - edgeSpectrum}(b)].

The characterization of the driven lattice in terms of winding numbers becomes apparent from
Fig.~\ref{fig - edgeSpectrum}(c-e), which shows quasienergy spectra of the continuously driven lattice
with momentum $k_{\|}$ along the periodic direction of a strip with armchair termination~\cite{Termination} 
for three distinct topologically non-trivial cases. Here each pair of levels
traversing the gap with opposite slope corresponds to one chiral edge mode appearing on opposite boundaries of
the strip. The insufficient information contained in Chern numbers $C_\pm=\pm(W_\pi-W_0)$
becomes apparent when considering the identical values ($|C_\pm|=1$) for the topologically distinct
cases shown in subfigures (c) and (d), and their trivial value ($C_\pm=0$) despite the topologically
non-trivial situation in subfigure (e).

\begin{figure}
	\centering\includegraphics[width=1\linewidth]{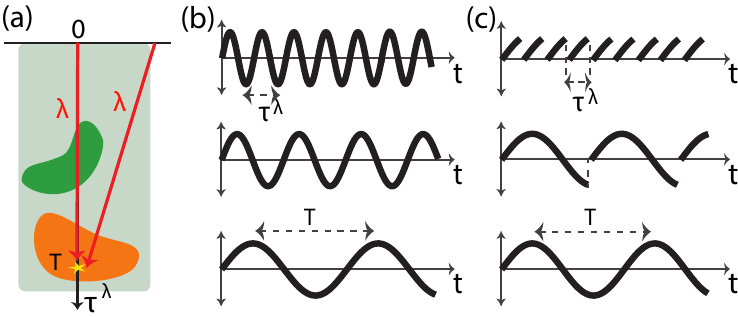}
	\caption{(a) General scheme of constructing the family of drives. Topologically distinct phases are depicted with different colors in the parameter space given by the drive period $\tau^\lambda$ and other parameters (symbolized by the horizontal axis). Two ways of reaching the target Hamiltonian with period $T$ [marked by a star in (a)] are: (b) increasing the driving period $\tau^\lambda$ and (c) chopping and repeating the drive at $\tau^\lambda$.}
    \label{fig - drive}
\end{figure}

\textit{Brute-force scheme for measuring $W_g$.}---%
In principle, $W_g$ may be measured by full tomographic reconstruction of the time-evolution operator $\mathcal{U}(\bk,t)$. In this brute-force scheme, one has to evolve the system
starting from a band-insulating state and perform full state tomography \cite{Hauke14_PRL, Flaschner16_Sci,Li16_Science} for a sufficiently dense time grid within the drive period. Moreover, once $\mathcal{U}(\bk,t)$ is obtained approximately, finding $W_g$ requires further
manipulations.
A smooth deformation $\mathcal{U}\to \mathcal{U}_g$ to a periodic operator $\mathcal{U}_g(t)=\mathcal{U}_g(t+T)$ is needed that leaves the gap $g$ open and makes all quasienergy bands degenerate at
$\pi-g$~\cite{Rudner13_PRX,YaoYanWan17a}. This is achieved, e.g., by the micromotion operator
$\mathcal{U}_g = \mathcal{U} \exp(i \mathcal{H}_F^g t)$, where $\mathcal{H}_F^g$ is a Floquet Hamiltonian
defined with the branch-cut of $\log$ along $e^{ig}$~\cite{Ug}
. Spatiotemporal derivatives of $\mathcal{U}_g(\bk,t)$ have to be estimated from the time and quasimomenta grids. Finally, the invariants can be computed from a discrete approximation of the three-dimensional integral of $W_g=\frac{1}{8\pi^2}\int\!\dd t \dd^2\bk\, \mathrm{tr}(\mathcal{X}^g_t [\mathcal{X}^g_{k_x},\mathcal{X}^g_{k_y}])$, with $\mathcal{X}^g_\eta = \mathcal{U}_g^{-1}\partial_\eta\mathcal{U}_g$. The complexity of calculating winding numbers by constructing the evolution operators renders this brute-force approach experimentally challenging. Let us, therefore, present another, more efficient scheme for measuring the winding numbers.

\textit{Efficient scheme.}---%
The main idea of this scheme is to start in the \emph{high-frequency regime}, which is topologically equivalent to a static system. We then identify a one-parameter family of drives that connects this static Hamiltonian to the target Hamiltonian, whose topology we want to probe. The quasienergy spectrum evolves as we tune through this family and topological transitions occur at quasienergy band touching points. These are topological singularities carrying a signed topological charge that defines the change in the topological invariants across the transition. Summing over these charges yields the topological invariant associated to each gap of the target Hamiltonian. Therefore, our general scheme consists of: (i) constructing the family of drives connecting the high-frequency regime to the target Hamiltonian; (ii) identifying the band touching points; and (iii) measuring the topological charge of these singularities individually via state tomography. Our scheme is motivated by the classification of Floquet topological insulators via phase bands~\cite{Nathan15_NJP}. However, while phase bands are an abstract theoretical concept, our families of drives can be implemented in concrete experiments. Let us now describe each step in detail.

\textit{Family of drives.}---%
We first identify the family of Hamiltonians $\hat H^\lambda(t)$ with period $\tau^{\lambda}$ by varying a (set of) variable(s), parameterized by $\lambda\in[0,1]$, that map to the target Hamiltonian $\hat H^1(t)=\hat H(t)$ with period $\tau^1=T$, see Fig.~\ref{fig - drive}(a). Different choices of $\hat H^\lambda$ take the system through different paths and different sets of topological phase transitions. The key ingredient is that all such paths start at a high-frequency regime with $\omega^\lambda = 2\pi/\tau^\lambda$ much larger than $|J(t)|$ and $|\Delta|$. Consequently, one can clearly identify the  $\pi$ gap as the larger gap, $\gamma_0\ll\gamma_\pi \sim \omega$, and conclude that the associated winding number vanishes, $W_\pi=0$. Thus, the topological classification of this initial driven system reduces to that in equilibrium~\cite{Nathan15_NJP}, where Chern numbers are sufficient for full characterization. Moreover, for our specific models, $W_0=0$ in the high-frequency regime since all coupling between the bands comes from higher-order corrections~\cite{Eckardt17_Rev}.

\begin{figure}
	\centering\includegraphics[width=1\linewidth]{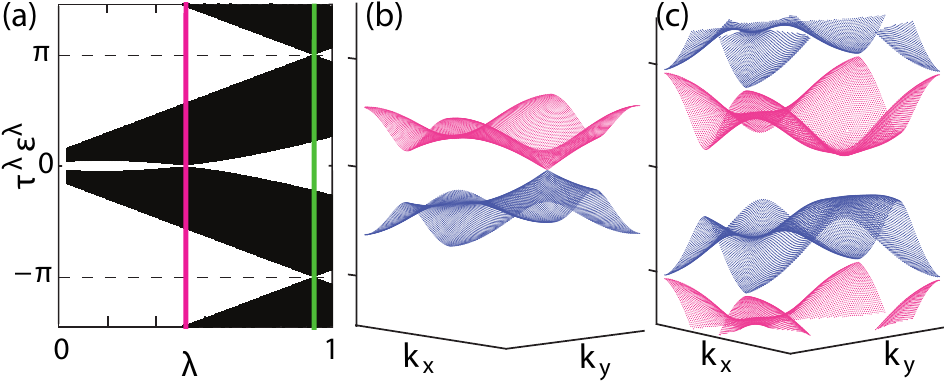}
	\caption{ Quasienergy bands of the circularly shaken honeycomb lattice as the drive frequency is lowered from $\omega=10J$ to target frequency $\omega=3J$ for $\kappa=1.5,\:\Delta=J$. (b,c) $\bk$-space plots at topological transitions marked by vertical lines in (a), with topological charges $q_0=q_\pi=1$. }
    \label{fig - quasienergy_lowerW}
\end{figure}

\begin{figure}
	\centering\includegraphics[width=1\linewidth]{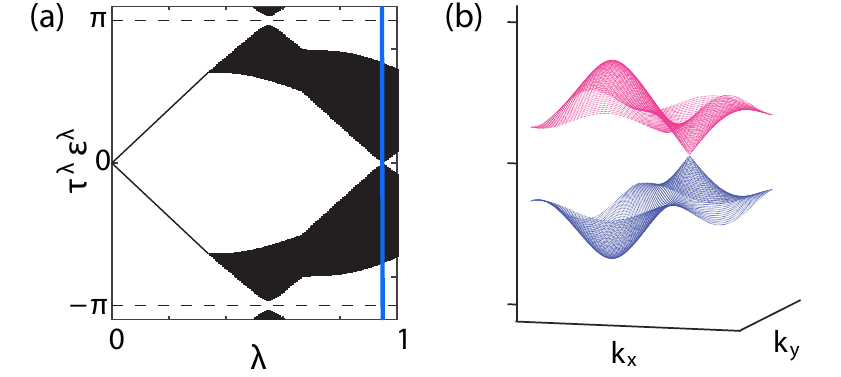}
	\caption{ Quasienergy bands of partial drives constructed for the step-wise drive with $\omega=3J$, and $\Delta=0$, with a topological charge $q_0=1$.}
    \label{fig - quasienergy_partial}
\end{figure}

The conditions on $\tau^{\lambda}$ are naturally satisfied if we tune the driving frequency monotonously while keeping the global shape of the drive fixed (~\ref{fig - drive}(b)), corresponding to a parametrization $\hat H^\lambda(t) = \hat H(t/\lambda)$.
Alternatively, we also consider a family of drives obtained by chopping the drive after a subperiod time $\tau^\lambda=\lambda T$ and repeat periodically thereafter. That is, $H^\lambda(t)=H(t \mod \tau^\lambda )$, as in Fig.~\ref{fig - drive}(c).

In Fig.~\ref{fig - quasienergy_lowerW}(a), we show the evolution of the bulk quasienergy spectrum of a continuously driven lattice as the period is increased up to the target value $T=2\pi/3J$ at fixed sublattice offset $\Delta=J$ and drive strength $\kappa=1.5$. The linear band touchings are zoomed in in Fig.~\ref{fig - quasienergy_lowerW}(b,c).
On the other hand, Fig.~\ref{fig - quasienergy_partial} demonstrates the evolution of the quasienergy bands for the drive-chopping scheme applied to step-wise drive with sublattice offset $\Delta=0$ and target period $T=2\pi/(3J)$. During the first stage of the drive, $\lambda<1/3$, the Hamiltonian is $\bk$-independent with flat quasienergy bands.

\begin{figure}
	\centering\includegraphics[width=1\linewidth]{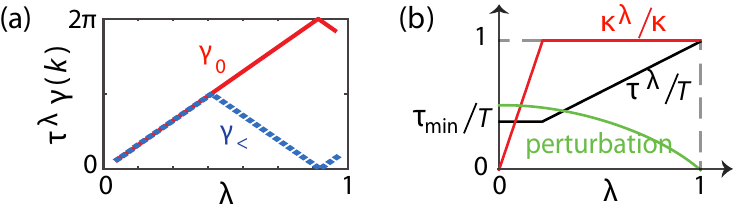}
	\caption{(a) The measured frequency $\gamma_<$ at $\bk=0$ and the quasienergy gap $\gamma_0$ for the singularity shown in Fig.~\ref{fig - quasienergy_lowerW}(c). The cusp indicates that the association of the $\gamma_{0}$ and the $\gamma_{\pi}$ gap as the smaller gap changes. (b) Experimental protocol starting from a high-frequency with a small period $\tau^0\ll |J|, |\Delta|$, the drive strength $\kappa^\lambda$ is ramped up to its final value and symmetry-breaking perturbation are smoothly turned off in the target Hamiltonian. }
    \label{fig - charge}
\end{figure}

\textit{Identifying singularities.}---%
We now introduce a protocol for detecting band-touching points of the quasienergy spectrum with respect to $\lambda$. For this purpose, one can perform a quench from an arbitrary band insulating state to the Hamiltonian $\hat H^{\lambda}(t)$. The stroboscopic evolution in steps of the driving period $\tau^\lambda$ is governed by the Floquet Hamiltonian $\hat{H}_F^\lambda$, which for each quasimomentum $\bk$
defines, a single angular frequency determined by $2|\bm h_F^\lambda(\bk)|$. Modulo $\omega^\lambda=2\pi/\tau^\lambda$ this frequency can be chosen to lie  in the interval $[-\omega^\lambda/2, \omega^\lambda/2]$, we can measure its absolute value $\gamma_<^{\lambda}(\bk)\in[0,\omega^\lambda/2]$ by monitoring the evolution of the stroboscopic momentum distribution at momentum $\bk$ ~\cite{Hauke14_PRL}. This frequency corresponds to the smaller one of the two band gaps $\gamma_0^{\lambda}(\bk)$ and $\gamma_\pi^{\lambda}(\bk)$, while the larger one is given by $\gamma_>^{\lambda}(\bk)=\omega^\lambda-\gamma_<^{\lambda}(\bk)$. In the high frequency limit, we can clearly identify $\gamma_<^{\lambda}(\bk)=\gamma_0^{\lambda}(\bk)$. Moreover, measuring $\gamma_<^{\lambda}(\bk)$ as a function of $\lambda$, one can also identify  when this association changes, i.e.\ when $\gamma_\pi^{\lambda}(\bk)$ becomes smaller than $\gamma_0^{\lambda}(\bk)$ so that  $\gamma_\pi^{\lambda}(\bk)=\gamma_<^{\lambda}(\bk)$. Namely, here $\gamma_<^{\lambda}(\bk)$ reaches a cusp-type maximum at $\omega^\lambda/2$, assuming that $2|\bm h^\lambda(\bk)|$ is a smooth function of $\bk$ [cf.\ Fig.~\ref{fig - charge}(a)]. Once the gaps $\gamma_g^{\lambda}$ are measured, one can identify the band touching singularities $\bm{p}^s$ in the three-dimensional parameter space $\bm{p}\equiv(\bk,\lambda)$.

\textit{Measuring topological charge.}---
Each singularity carries a topological charge determined by the winding of
$\bm{h}_F(\bm{p}) \equiv\bm{h}_F^\lambda(\bm{k})$ around $\bm{p}^s$ in the three-dimensional manifold~\cite{Nathan15_NJP}.
For a linear $g$-gap closing, $W_g$ changes by $q_s = \sgn(\det S) = \pm1$, where $S_{ij}=\partial h_{F i}/{\partial p_j}\vert_{\vex p^s}$. We now describe how the matrix $S$ and the topological charge can be measured experimentally.

To construct $S$, we need the Floquet Hamiltonian at six points, in principle, infinitesimally away from the singularity at $\vex p^s$. However, since the topological charge is quantized, therefore, protected against small perturbations, we can consider the Floquet Hamiltonian at finite distances $\Delta\vex p$. We find that the charge $q_s$ is extremely robust against the distance $\Delta\vex p$ away from a singularity. Deviations occur only when $\Delta k$ approaches halfway across the BZ, and similar rigidity is observed along $\Delta\lambda$ as evident from the linear profile of the quasienergy in Figs.~\ref{fig - quasienergy_lowerW} and \ref{fig - quasienergy_partial}.
While the bands are degenerate at the singularity, they are gapped around it. This allows for an adiabatic state preparation at the six tomography points. Namely, by choosing these points to have a finite distance from $\bm{p}^s$ in quasimomentum (e.g.\ by displacing them in diagonal $\bm{p}$-directions from the singularity), they can be reached without gap closing by ramping $\lambda$. Moreover, the small-$\lambda$ (high-frequency) regime can be reached from undriven system without crossing a phase transition by smoothly ramping up the driving amplitude.
The states so prepared, which contain all the information of the Floquet Hamiltonian at the tomography point,
can then be measured via state tomography by suddenly projecting the system onto flat bands~\cite{Hauke14_PRL,Flaschner16_Sci}, from which the matrix $S$ and the topological charge of the singularity can be obtained. Once the charges of all the singularities are determined, the topological invariant of each gap $g$ of the target Hamiltonian is obtained by summing over the charges $q^s_g$ of the singularities encountered for this gap, while tuning $\lambda$ from 0 to 1, $W_g = W_g^0 + \sum_{s_g} q_{s_g}$. Here $W_g^0$ is the value of the invariant in the high-frequency regime: $W_0^0$ equals the Chern number of the high-frequency Hamiltonian, whereas $W_\pi^0=0$.

\textit{Discussion.}---%
Some band degeneracies might be extended in the parameter space $\vex p$, due to accidental symmetries~\cite{Herring37_PhysRev_accDegeneracy}. While these extended degeneracies do not pose a problem for the detection of the quasienergy spectrum, they complicate the measurement of topological charges. Fortunately, they can be resolved by adding small symmetry-breaking perturbations as depicted in Fig.~\ref{fig - charge}(b): if indeed there is a singularity hidden in the extended degeneracy, adding such a perturbation would only shift its position in $\vex p$ and preserve its topological charge. For example, in Fig.~\ref{fig - quasienergy_partial}(a) there was originally an extended $\gamma_\pi$-gap closing, which was resolved by anisotropically reducing the tunnel coupling in one direction by $20\%$.

Several remarks are in order regarding the family of drives (cf.\ Fig.~\ref{fig - drive}). In any realistic experiment, infinite-frequency $\tau^0\to0$ cannot be attained. In practice, it is sufficient to start from a frequency much larger than $|J|$ and $|\Delta|$, e.g.\ $\tau^0=2\pi/10J$ in Fig.~\ref{fig - quasienergy_lowerW}. We sketch such a realistic experimental protocol in Figure \ref{fig - charge}(b) where in a first step the driving amplitude is ramped from zero to its final value. Also, in the partial drive scheme [cf.\ Fig.(c)], a net force may be induced on the atomic cloud in the optical lattice. In order to prevent atom loss in this case, one can modify the chopped drive by adding also a constant counter-reacting force to cancel the drift introduced by the truncation.

In the chopped-drive scheme [cf.\ Fig.~\ref{fig - drive}(c)], the one-cycle evolution operator of the constructed system Hamiltonian directly corresponds to the evolution operator $\Uo(\tau)$ of the target system, with $\tau<T$. Thus, the eigenstates of $\Uo(\tau)$ are turned into Floquet modes and the phases of its eigenvalues (corresponding to the phase bands introduced in Ref.~\cite{Nathan15_NJP}) into quasienergies. This allows one to measure the eigenvalues of $\Uo(\tau)$ from the time evolution and to prepare eigenstates of $\Uo(\tau)$ adiabatically. This chopping and repeating of the Hamiltonian in time can be viewed as a time analog of crystallography. Instead of periodically repeating a spatial structure (i.e.\ crystallizing a complex molecule) for measuring it via diffraction, a temporal structure is repeated for probing its properties. This concept can be generalized to any time evolution generated by  a time-dependent Hamiltonian, periodic or not.

Finally, we would like to stress that our scheme does not require the adiabatic preparation of the full topological band insulator, but rather relies on the tomography at a number of $k$-space points only, making it robust and efficient (an alternative and completely different strategy for measuring $W_g$ without full state preparation can be based on observing post-quench dynamics within single driving periods~\cite{Unal19_arx}). Our approach can be a useful tool also for the characterization of topological insulators that are equivalent to static systems, for which $W_\pi=0$. 
Since the relationship between winding numbers and topological charges of band-touching singularities is general, our scheme would also work for systems with more than two bands by extending the tomography techniques suitably.

\begin{acknowledgments}
	This work was supported in part by the National Science Foundation CAREER award DMR-1350663, the US-Israel Binational Science Foundation under grant No. 2014245, the Deutsche Forschungsgemeinschaft via the Research Unit FOR 2414 (Project No.~277974659), and the College of Arts and Sciences at Indiana University. The authors acknowledge fruitful discussions with Robert-Jan Slager and Christof Weitenberg.
\end{acknowledgments}

\bibliography{references}

\begin{thebibliography}{49}
\bibitem [{Ter()}]{Termination}%
  \BibitemOpen
  \href@noop {} {\enquote {\bibinfo {title} {For a different termination, say
  zig-zag, the spectra look different in their details but the three distinct
  phases can still be identified in the same parameter regimes showing the same
  number of chiral edge modes.}}\ }\BibitemShut {NoStop}%
\bibitem [{Ug()}]{Ug}%
  \BibitemOpen
  \href@noop {} {\enquote {\bibinfo {title} {This choice of $\mathcal{U}_g$ as
  a micromotion operator transparently shows how the topological information is
  split between the stroboscopic evolution ($\mathcal{H}_f^g$) and the
  micromotion: the chern numbers of the bands of $\mathcal{H}_f^g$ together
  with one winding number $w_g$ (obtained from $\mathcal{U}_g$) provide a full
  topological characterization of the driven lattice.}}\ }\BibitemShut
  {NoStop}%
\end{thebibliography}

\end{document}